\documentclass[final,1p,times]{elsarticle} 

\usepackage{graphicx}
\usepackage{amssymb} 
\usepackage{amsthm} 
\usepackage{lineno}
\usepackage{color}

\journal{Nuclear Physics A} 

\begin{document}

\begin{frontmatter} 

\title{Net-­‐charge fluctuations and balance functions at the LHC}

\author{M.~Weber, for the ALICE\fnref{col1} Collaboration}
\fntext[col1] {A list of members of the ALICE Collaboration and acknowledgements can be found at the end of this issue.}
\address{University of Houston, Physics Department, Houston, TX 77204}
\ead{m.weber@cern.ch}


\begin{abstract}
The measurement of event--by--event fluctuations and charge--dependent particle correlations are used to study properties of nuclear matter at high temperatures as produced in ultrarelativistic heavy--ion collisions. We present results for event--by--event net--charge fluctuations and charge balance functions in $\Delta\eta$ and $\Delta\varphi$ in Pb--Pb collisions at $\sqrt{s_{\mathrm {NN}}}= 2.76$ TeV. 

\end{abstract}

\end{frontmatter} 




\section{Motivation}
\label{sec:motivation}

The study of strongly interacting matter at extreme energy densities as produced in ultrarelativistic heavy--ion collisions is the aim of the ALICE detector \cite{Ref:Alice} at the Large Hadron Collider (LHC) at CERN. Quantum chromodynamics (QCD) \cite{Ref:QCD} predicts that at sufficiently high energy density, of the order of 0.5~GeV/fm$^3$ \cite{Ref:QGP}, a deconfined state of quarks and gluons is produced, the so-called quark gluon plasma (QGP). Many different observables are used to find experimental evidence and subsequently study the properties of the QGP.  In this contribution we will focus on the measurement of event--by--event net--charge fluctuations and charge balance functions. The analyses use the data taken in Pb--Pb collisions at $\sqrt{s_{\mathrm {NN}}}= 2.76$ TeV and make use of the excellent tracking capabilities of the ALICE Time Projection Chamber (TPC). 

\section{Net--charge fluctuations}
\label{sec:netchargefluctuations}

Insight into the possible existence of a QGP can be gained by determining the relevant degrees of freedom for the electric charge of the system produced in a heavy--ion collision \cite{Ref:NetChargeFluctuations}. In a hadron (resonance) gas these are mesons and baryons, while in the QGP phase these would be quarks in the simplest case (neglecting more complex bound states). 
A significant difference between the two phases is expected in measuring event--by--event net--charge fluctuations per charged degree of freedom:
\begin{equation}
D=4\frac{\langle\delta Q^2\rangle}{\langle N_{\mathrm{ch}}\rangle} \approx \langle N_{\mathrm{ch}} \rangle\nu_{+-,\mathrm{dyn}} + 4
\label{Equ:D}
\end{equation}
with $\langle\delta Q^2\rangle$ the variance of the net--charge $Q=N_{+}-N_{-}$ and $\langle N_{ch} \rangle=\langle N_{+}+N_{-} \rangle$ the average number of all measured charged particles per event. The variable D is close to 1 for the case of a QGP, and is predicted to be close to 3 for a hadron (resonance) gas \cite{Ref:chargeFluctuationsTheory}. 

In the experiment, the net--charge fluctuations are best studied by calculating the quantity $\nu_{+-,dyn}$, which is strongly connected to the variable D via the relationship shown in Eq.~\ref{Equ:D} and is defined as:
\begin{equation}
\nu_{+-,\mathrm{dyn}}=\frac{\langle N_+(N_+-1)\rangle}{\langle N_+\rangle^2}+\frac{\langle N_-(N_--1)\rangle}{\langle N_-\rangle^2}-2\frac{\langle N_-N_+\rangle}{\langle N_-\rangle\langle N_+\rangle}
\label{Equ:nu}
\end{equation}
Taking into account global charge conservation and finite acceptance, one obtains the corrected value of $\nu_{+-,\mathrm{dyn}}$:
\begin{equation}
\nu_{+-,\mathrm{dyn}}^{\mathrm{corr}}=\nu_{+-,\mathrm{dyn}} + \frac {4}{ \langle N_{\mathrm{total}}\rangle}
\label{Equ:nucorr}
\end{equation}
where $\langle N_{\mathrm{total}}\rangle$ is the average total number of charged particles produced over the full phase space.

\begin{figure*}[htb]
\begin{minipage}[b]{0.52\linewidth}
\includegraphics[width=\linewidth]{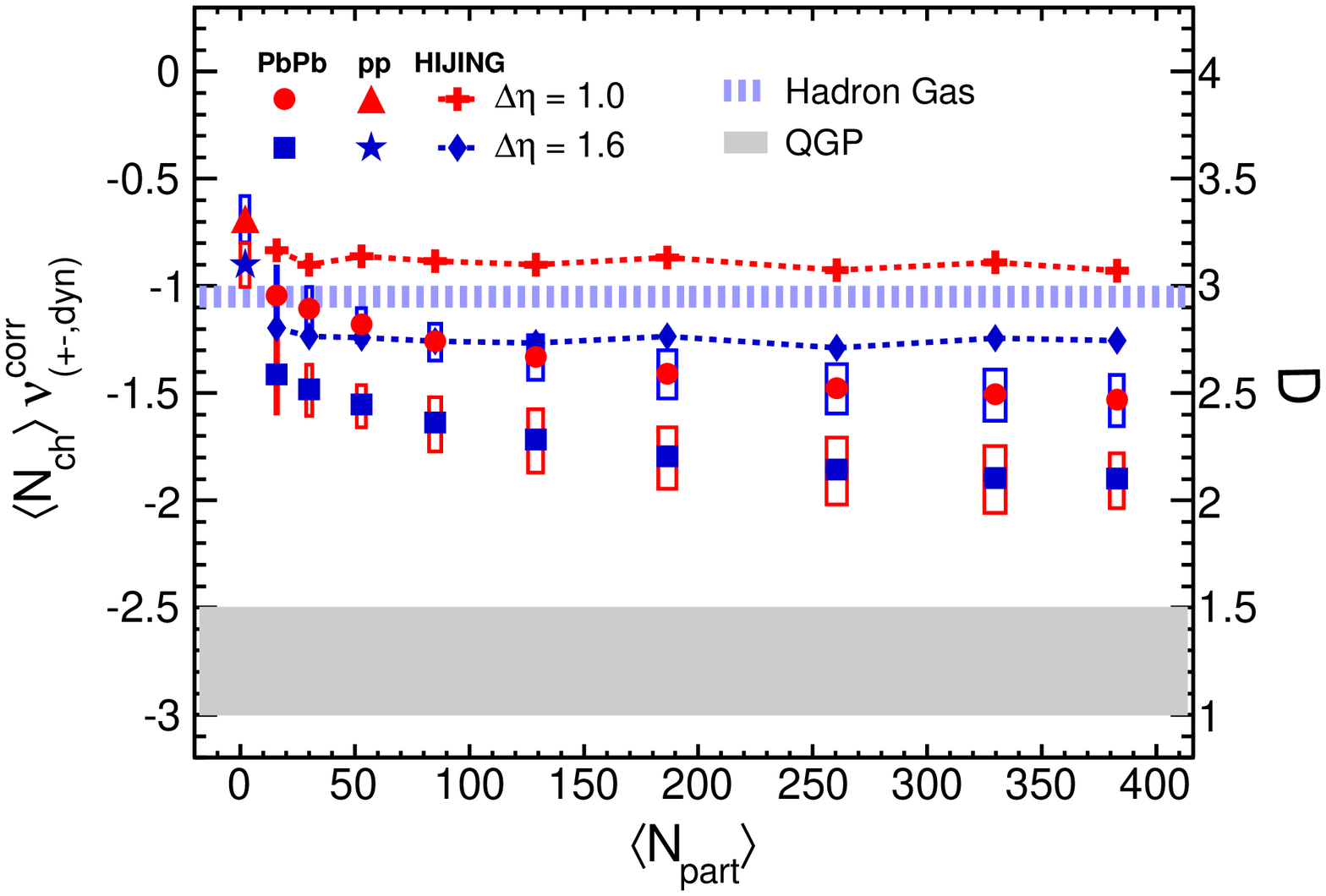}
\end{minipage}
\begin{minipage}[b]{0.52\linewidth}
\includegraphics[width=\linewidth]{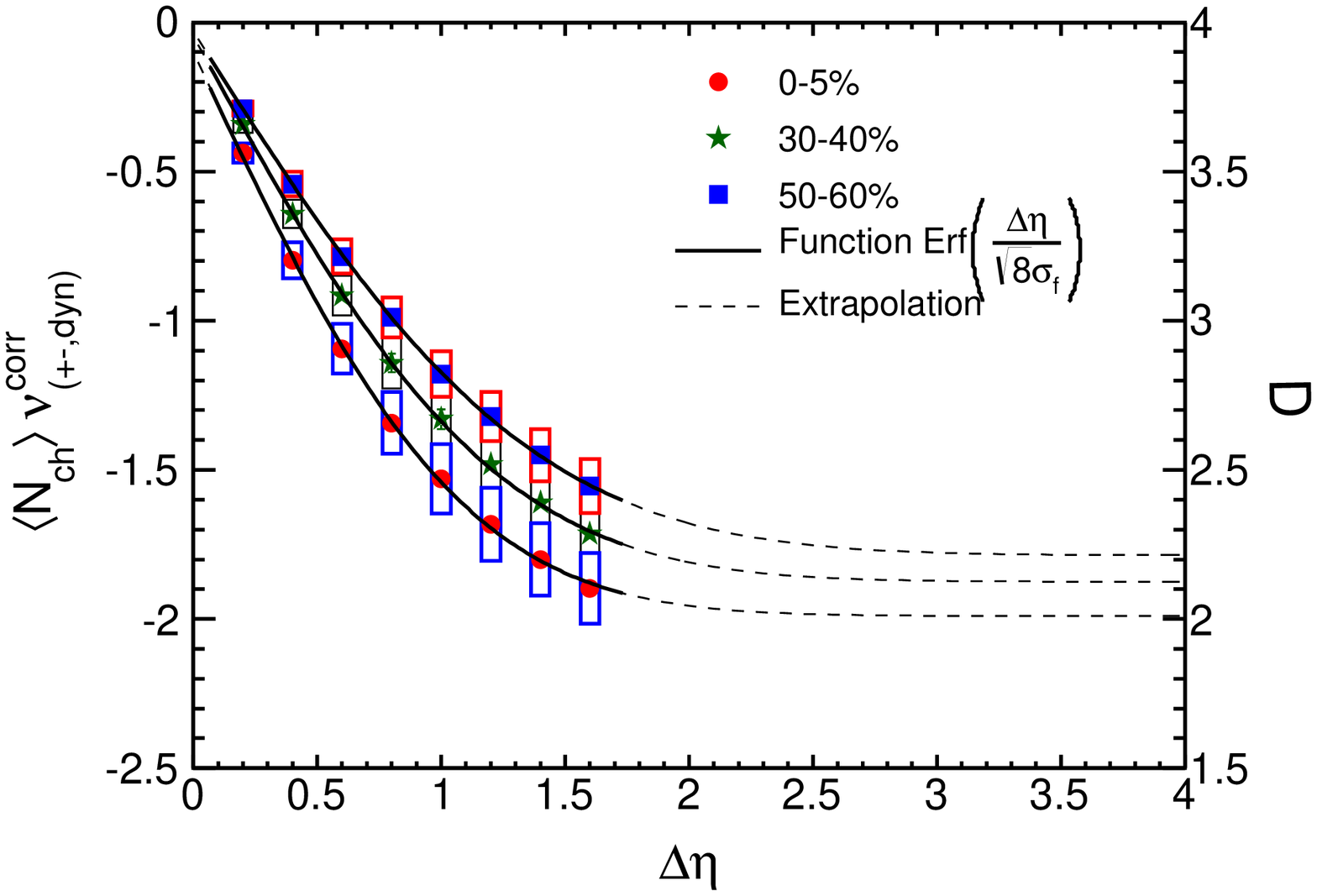}
\end{minipage}
\caption{(color online) Left: $\langle N_{\mathrm{ch}} \rangle\nu_{+-,\mathrm{dyn}}^{\mathrm{corr}}$ (left axis) and $D$ (right axis) as a function of the number of participants for $\Delta\eta=1$ and $\Delta\eta=1.6$ in Pb--Pb collisions at $\sqrt{s_{\mathrm {NN}}}=2.76$~TeV. Also shown are the results from the HIJING event generator for both the $\Delta\eta$ windows and in the shaded bands the expectations for a hadron resonance gas and the QGP \cite{Ref:chargeFluctuationsTheory}. Right: $\langle N_{\mathrm{ch}} \rangle\nu_{+-,\mathrm{dyn}}^{\mathrm{corr}}$ (left axis) and $D$ (right axis) as a function of $\Delta\eta$ for different centrality percentiles. The data points are fitted with the functional form, $erf(\frac{\Delta\eta}{\sqrt{8}\sigma_{\mathrm f}})$. The dashed lines correspond to the extrapolation of the fitted curves. Both statistical (error bar) and systematic (box) errors are shown. Figures from \cite{Ref:NetChargeFluctuations}.}
\label{fig:chargeFluc:nudyn}
\end{figure*}

In the left panel of Fig.~\ref{fig:chargeFluc:nudyn} the values of $\langle N_{\mathrm{ch}} \rangle\nu_{+-,\mathrm{dyn}}^{\mathrm{corr}}$ are shown for pp and Pb--Pb collisions as a function of the mean number of participating nucleons  $\langle N_{\mathrm{part}} \rangle$ in two pseudorapidity windows $\Delta\eta = 1$ and $\Delta\eta = 1.6$. The shaded bands indicate the predictions for a hadron resonance gas (HG) and the QGP \cite{Ref:chargeFluctuationsTheory}. The data points show a monotonic decreasing dependence with $\langle N_{\mathrm{part}} \rangle$ and lie clearly below the HG, but above the QGP expectation. On the other hand, the results from  the HIJING event generator for both pseudorapidity windows do not show any dependence on $N_{\mathrm{part}}$ and are in the vicinity of the HG band.  The observed pseudorapidity dependence, as shown in the right panel of Fig.~\ref{fig:chargeFluc:nudyn}, hints to a dilution of the primordial fluctuations during the evolution of the system from hadronization to kinetic freeze--out because of the diffusion of charged hadrons in rapidity \cite{Ref:FlucDiffusion}. It can be parametrized with 
\begin{equation}
erf(\frac{\Delta\eta}{\sqrt{8}\sigma_{\mathrm f}})
\end{equation}
with $\sigma_{\mathrm f}$ being a measure for the diffusion.

\newpage

\section{Balance functions}
\label{sec:balancefunctions}

\begin{figure*}[htb]
\centering
\includegraphics[width=\linewidth]{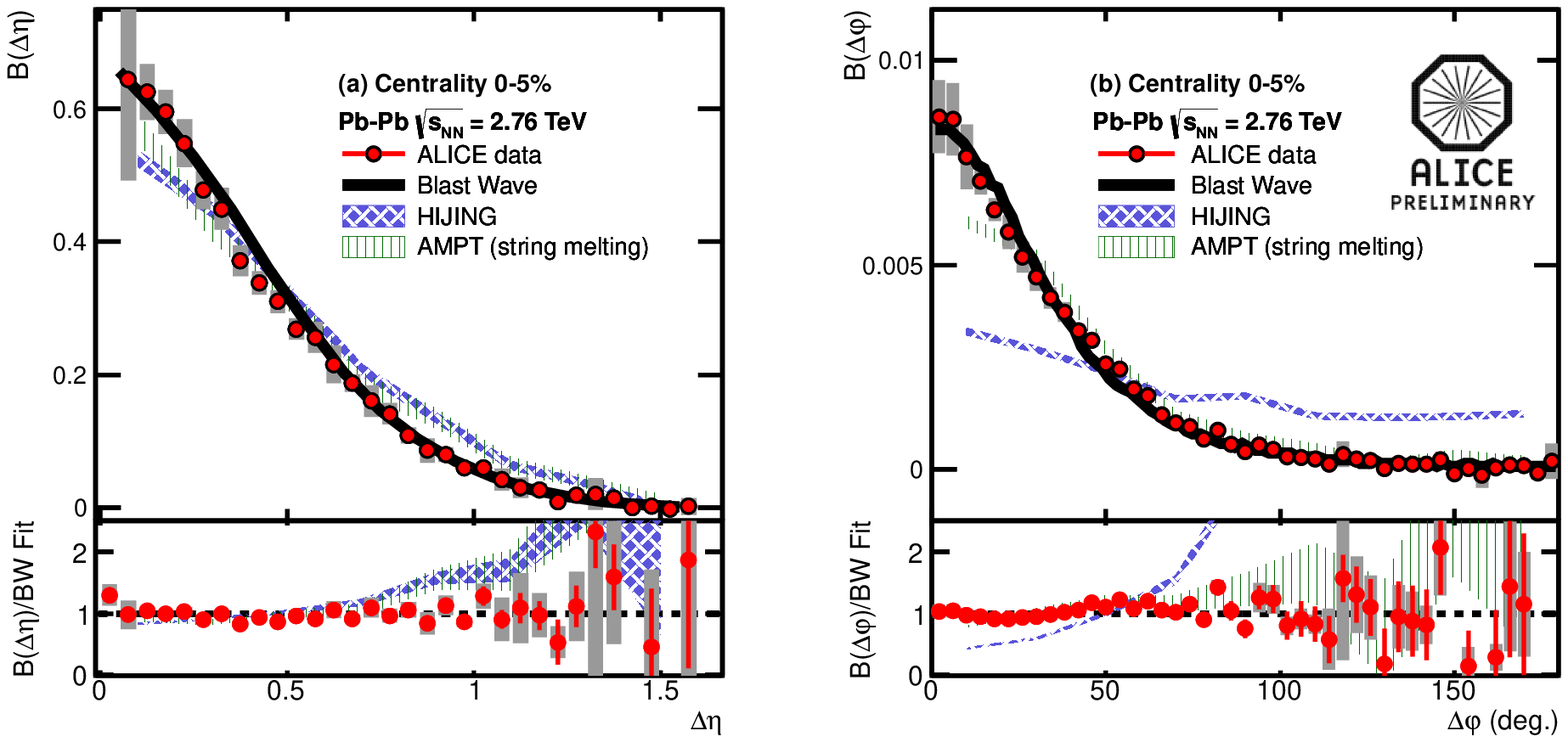}
\includegraphics[width=\linewidth]{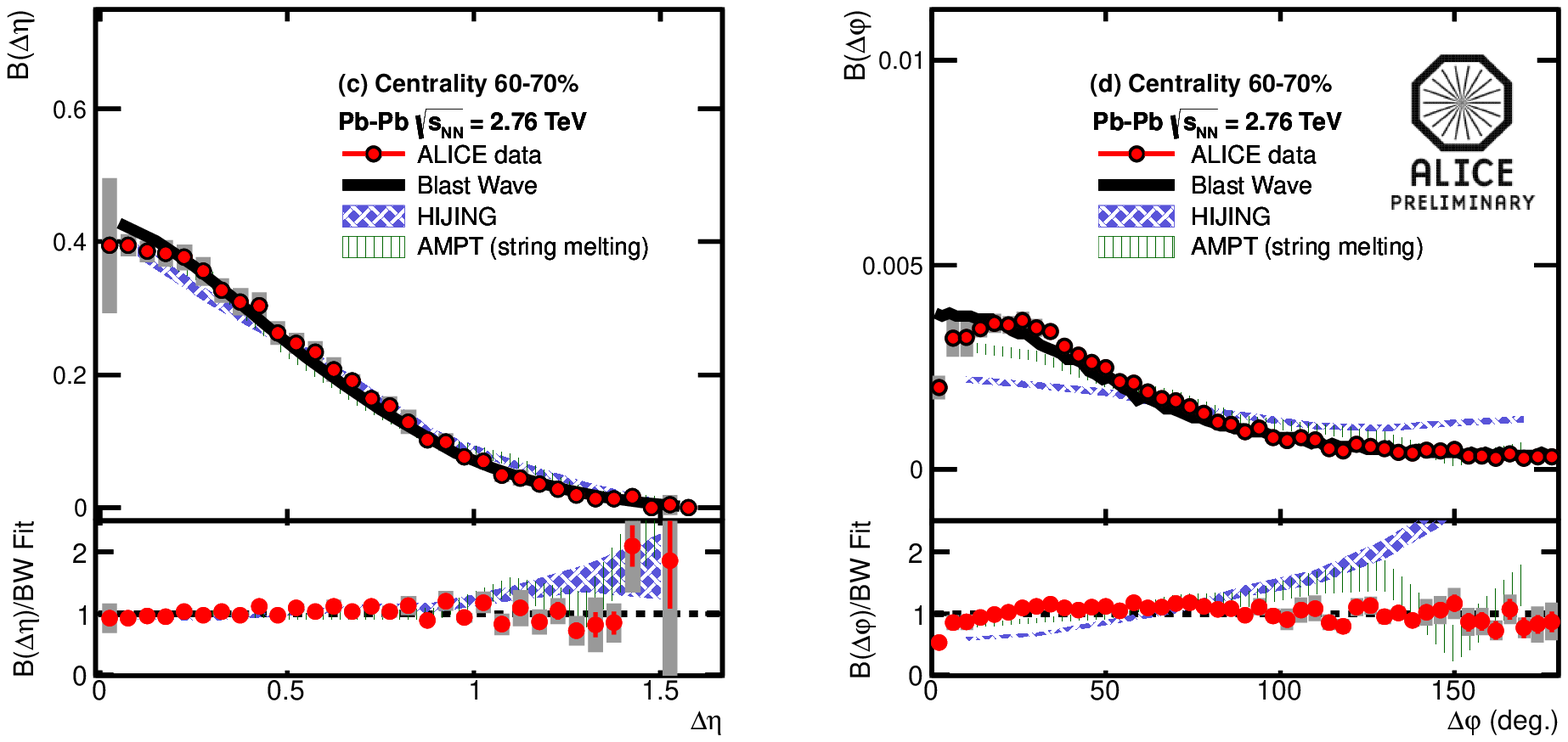}
\caption{(color online) The balance function distributions for the 5\% most central (a,b) and peripheral (c,d) Pb--Pb collisions at $\sqrt{s_{\mathrm {NN}}}=2.76$~TeV \cite{Ref:ALICEBF} as a function of the relative pseudorapidity $\Delta\eta$ (a,c) and the relative azimuthal angle $\Delta\varphi$ (b,d). The data points are compared to results from HIJING, from AMPT (string melting) and from a thermal blast wave. The ratios to the blast wave fit are shown in the lower panels. The models are normalized to the same integral as the ALICE data.}
\label{fig:BF}
\end{figure*}

It was proposed to measure the creation time of particles via the correlations between positive and negatively charged pairs as a function of rapidity \cite{Ref:BF_theory}. Assuming these pairs being created at the same space--time point and correlated in momentum space due to a strong collective expansion, their separation in pseudorapidity $\Delta\eta$ depends not only on the initial momentum difference but also on the length of the rescattering phase. Furthermore, it was shown that the balance function for the relative azimuthal angle of the charge--anticharge pair can probe the collective motion of the produced system and in particular its radial flow \cite{Ref:BF_phi}.

The definition of the balance function (e.g. for the pseudorapidity difference $\Delta\eta$) reads:
\begin{equation}
B(\Delta \eta) = \frac{1}{2} \Big[ \frac{N_{+-}(\Delta \eta) -
N_{--}(\Delta \eta)}{N_{-}} + \frac{N_{-+}(\Delta \eta) -
N_{++}(\Delta \eta)}{N_{+}}  \Big]
\label{Eq:BFDefinitionInEta}
\end{equation}
with $N_{--(++,+-)}(\Delta \eta)$ the number of $ -- (++, +- )$ particle pairs and $N_{+(-)}$ the number of positive (negative) particles in the analyzed phase--space. 

In Fig.~\ref{fig:BF} the balance functions are shown for the 5\% most central and for more peripheral (60--70\%) Pb--Pb collisions as a function of $\Delta\eta$ and $\Delta\varphi$ \cite{Ref:ALICEBF}. Furthermore, a comparison to results from different MC event generators is shown. HIJING \cite{Ref:HIJING} produces charges early in the collision history, mainly via string fragmentation, and includes no collective motion, e.g. radial flow. It is not able to match the data points for $B(\Delta \eta)$ and $B(\Delta \varphi)$ for central collisions. Only in the most peripheral events the HIJING results for $B(\Delta \eta)$ are consistent with the ALICE data, which shows a strong centrality dependence, that is not observed in HIJING.
AMPT (string melting) \cite{Ref:AMPT}, on the other hand, with parameters tuned to reproduce the measured elliptic flow values of non--identified particles at the LHC, shows an agreement in $B(\Delta \varphi)$, not only in the most central bin but for the whole centrality dependence. This can be understood as collective flow being the determining source of balancing charge correlation in $\Delta\varphi$. For $B(\Delta \eta)$ the balance functions are similar to HIJING.

\section{Summary and outlook} 
\label{sec:summaryandoutlook}

In summary, dynamical net--charge fluctuations were presented as a function of centrality and pseudorapidity and have a value below the expectation for a hadron resonance gas and above that of the QGP. The balance functions for charged particles show a strong narrowing with increasing centrality, which can be reproduced in $\Delta\varphi$ with a model including collective effects, e.g. flow, but not in $\Delta\eta$. The analysis of net--charge fluctuations will be extended to higher moments and balance functions will be studied for identified particles and regarding their transverse momentum and event--plane dependence.







\begin{thebibliography}{00}


\bibitem{Ref:Alice} {
K.~Aamodt \textit{et al.} [ALICE Collaboration], J. Phys. \textbf{G30}, (2004) 1517; \\
K.~Aamodt \textit{et al.} [ALICE Collaboration], J. Phys. \textbf{G32}, (2006) 1295.\\
K. Aamodt \textit{et al.} [ALICE Collaboration], JINST \textbf{3}, (2008) S08002.
}

\bibitem{Ref:QCD} {H.~Satz, Rep. Prog. Phys. \textbf{63}, (2000) 1511; \\
S.A.~Bass, M.~Gyulassy, H.~St\"{o}cker, W.~Greiner, J. Phys. \textbf{G25}, (1999) R1; \\
E.V.~Shuryak, Phys. Rep. \textbf{115}, (1984) 151; \\
J.~Cleymans, R.V.~Gavai, E.~Suhonen, Phys. Rep. \textbf{130}, (1986) 217.}

\bibitem{Ref:QGP} {S.~Borsanyi, JHEP \textbf{1011}, (2010) 077.}
%
%
%


\bibitem{Ref:NetChargeFluctuations} {
B.~Abelev \textit{et al.} [ALICE Collaboration], \textit{arXiv:}1207:6068.
}

\bibitem{Ref:chargeFluctuationsTheory}{
S.~Jeon, V.~Koch, Phys. Rev. Lett. 85 (2000), 2076.\\
S.~Jeon and V.~Koch, In Quark–Gluon–
Plasma 3, Ed. R.C. Hwa and X.N. Wang, 430 (2004);
arXiv:hep-ph/0304012v1.
}

\bibitem{Ref:FlucDiffusion}{
E. V. Shuryak, M. A. Stephanov, Phys. Rev. C \textbf{63}, (2001) 064903.\\
M. A. Aziz, S. Gavin, Phys. Rev. C \textbf{70}, (2004) 034905.
}

\bibitem{Ref:BF_theory}{
S. Bass, P. Danielewicz, and S. Pratt, Phys. Rev. Lett. 85, (2000) 2689.\\
S. Jeon and S. Pratt, Phys. Rev. C65, (2002) 044902.
}	

\bibitem{Ref:BF_phi}{
P. Bozek, Phys. Lett. B609. (2005) 247. 
}	


\bibitem{Ref:ALICEBF} {
B.~Abelev \textit{et al.} [ALICE Collaboration], to be published.
}


\bibitem{Ref:HIJING}{
M.~Gyulassy and X.~N.~Wang, Comput. Phys. Commun. \textbf{83}, 307 (1994). \\
X.~N.~Wang and M.~Gyulassy, Phys. Rev. \textbf{D44}, 3501 (1991).
}

\bibitem{Ref:AMPT}{
B. Zhang et al., Phys. Rev. C61, (2000) 067901.\\ 
Z. W. Lin et al., Phys.Rev. C64 (2001) 011902. \\
Z. W. Lin et al., Phys. Rev. C72, (2005) 064901.
}



\end{thebibliography}



\end{document}